\documentclass[runningheads]{llncs}

\usepackage{graphicx}
\usepackage{xspace}
\usepackage{caption,subcaption}
\usepackage{tikz}
\usetikzlibrary{arrows}
\usepackage[noend]{algorithm2e}
\usepackage{amsmath}
\usepackage{colortbl}
\usepackage{stmaryrd}
\usepackage{hyperref}

\newcommand{\codeintext}[1]{\texttt{#1}}

\usepackage{listings, xcolor}

\definecolor{verylightgray}{rgb}{.97,.97,.97}
\definecolor{darkgreen}{rgb}{0,.5,0}

\lstdefinelanguage{Solidity}{
  basicstyle=\scriptsize\ttfamily, 
	keywords=[1]{anonymous, assembly, assert, balance, break, call, callcode, case, catch, class, constant, continue, contract, debugger, default, delegatecall, delete, do, else, emit, event, export, external, false, finally, for, function, gas, if, implements, import, in, indexed, instanceof, interface, internal, is, length, library, log0, log1, log2, log3, log4, memory, modifier, new, payable, pragma, private, protected, public, pure, push, require, return, returns, revert, selfdestruct, send, storage, struct, suicide, super, switch, then, this, throw, transfer, true, try, typeof, using, view, while, with, addmod, ecrecover, keccak256, mulmod, ripemd160, sha256, sha3}, 
	keywordstyle=[1]\color{blue}\bfseries,
	keywords=[2]{address, bool, byte, bytes, bytes1, bytes2, bytes3, bytes4, bytes5, bytes6, bytes7, bytes8, bytes9, bytes10, bytes11, bytes12, bytes13, bytes14, bytes15, bytes16, bytes17, bytes18, bytes19, bytes20, bytes21, bytes22, bytes23, bytes24, bytes25, bytes26, bytes27, bytes28, bytes29, bytes30, bytes31, bytes32, enum, int, int8, int16, int24, int32, int40, int48, int56, int64, int72, int80, int88, int96, int104, int112, int120, int128, int136, int144, int152, int160, int168, int176, int184, int192, int200, int208, int216, int224, int232, int240, int248, int256, mapping, string, uint, uint8, uint16, uint24, uint32, uint40, uint48, uint56, uint64, uint72, uint80, uint88, uint96, uint104, uint112, uint120, uint128, uint136, uint144, uint152, uint160, uint168, uint176, uint184, uint192, uint200, uint208, uint216, uint224, uint232, uint240, uint248, uint256, var, void, ether, finney, szabo, wei, days, hours, minutes, seconds, weeks, years},	
	keywordstyle=[2]\color{teal}\bfseries,
	keywords=[3]{block, blockhash, coinbase, difficulty, gaslimit, number, timestamp, msg, gas, sender, sig, value, now, tx, gasprice, origin},	
	keywordstyle=[3]\color{violet}\bfseries,
	identifierstyle=\color{black},
	sensitive=false,
	comment=[l]{//},
	morecomment=[s]{/*}{*/},
	commentstyle=\color{darkgreen}\ttfamily,
	stringstyle=\color{red}\ttfamily,
	morestring=[b]',
	morestring=[b]",
	numberstyle=\scriptsize,
	emphstyle=\underline
}

\lstdefinelanguage{Boogie}{
  basicstyle=\scriptsize\ttfamily, 
	keywords=[1]{axiom, break, call, const, else, exists, extends, forall, function, goto, if, implementation, modifies, old, procedure, returns, then, type, unique, var, while}, 
	keywordstyle=[1]\color{blue}\bfseries,
	keywords=[2]{address, bool, int, real}, 
	keywordstyle=[2]\color{teal}\bfseries,
	keywords=[3]{assert, assume, ensures, invariant, requires}, 
	keywordstyle=[3]\color{darkgreen}\bfseries,
	keywords=[4]{\_msg\_sender, \_msg\_value, \_balance, \_this}, 
	keywordstyle=[4]\color{violet}\bfseries,
	identifierstyle=\color{black},
	sensitive=false,
	comment=[l]{//},
	morecomment=[s]{/*}{*/},
	commentstyle=\color{gray}\ttfamily,
	stringstyle=\color{red}\ttfamily,
	morestring=[b]',
	morestring=[b]",
  numberstyle=\scriptsize
}

\newcommand{\solcverify}{\textsc{solc-verify}\xspace}
\newcommand{\myth}{\textsc{mythril}\xspace}
\newcommand{\verisol}{\textsc{verisol}\xspace}
\newcommand{\smtchecker}{\textsc{smt-checker}\xspace}

\newcommand{\cvc}{\textsc{cvc4}\xspace}
\newcommand{\zzz}{\textsc{z3}\xspace}

\newcommand{\tablevspace}{\vspace{0.5em}}

\newcommand{\soltypnam}{\textit{TypeName}}
\newcommand{\soladdrtyp}{\codeintext{address}}
\newcommand{\solinttyp}{\codeintext{int}}
\newcommand{\soluinttyp}{\codeintext{uint}}
\newcommand{\solbooltyp}{\codeintext{bool}}
\newcommand{\solmaptyp}[2]{\codeintext{mapping(}#1 \codeintext{=>} #2\codeintext{)}}
\newcommand{\soldynarr}[1]{#1\codeintext{[]}}
\newcommand{\solfixarr}[2]{#1\codeintext{[\ensuremath{#2}]}}
\newcommand{\sollval}{\textit{lval}}
\newcommand{\solexpr}{\textit{expr}}
\newcommand{\solstmt}{\textit{stmt}}

\newcommand{\smtint}{\textit{int}}
\newcommand{\smtbool}{\textit{bool}}
\newcommand{\smtarrtyp}[2]{\ensuremath{[#1]#2}}
\newcommand{\smtarrheap}[1]{\ensuremath{\mathit{arrheap}_{#1}}}
\newcommand{\smtstorarrtype}[1]{\ensuremath{\textit{StorArr}_{#1}}}
\newcommand{\smtmemarrtype}[1]{\ensuremath{\textit{MemArr}_{#1}}}
\newcommand{\smtstructheap}[1]{\ensuremath{\mathit{structheap}_{#1}}}
\newcommand{\smtstorstructtype}[1]{\ensuremath{\textit{StorStruct}_{#1}}}
\newcommand{\smtmemstructtype}[1]{\ensuremath{\textit{MemStruct}_{#1}}}
\newcommand{\smtconstarr}[3]{\ensuremath{\textsf{constarr}_{[#1]#2}(#3)}}

\newcommand{\mappedto}{\ensuremath{\dot{=}}\xspace}
\newcommand{\extradecl}[1]{[#1]}
\newcommand{\extrastmt}[1]{\{#1\}}
\newcommand{\maptype}[1]{\ensuremath{\mathcal{T}(#1)}}
\newcommand{\typeof}[1]{\ensuremath{\textsf{type}(#1)}}
\newcommand{\elementtypeof}[1]{\ensuremath{\textsf{arrtype}(#1)}}
\newcommand{\defval}[1]{\ensuremath{\textsf{defval}(#1)}}
\newcommand{\storagetree}{\textsf{tree}}
\newcommand{\unpackptr}[1]{\ensuremath{\textsf{unpack}(#1)}}
\newcommand{\packpath}{\ensuremath{\textsf{packpath}}\xspace}
\newcommand{\packptr}[1]{\ensuremath{\textsf{pack}(#1)}}
\newcommand{\mapexpr}[1]{\ensuremath{\mathcal{E}(#1)}}
\newcommand{\mapstmt}[1]{\ensuremath{\mathcal{S}\llbracket #1 \rrbracket}}
\newcommand{\mapassign}[2]{\ensuremath{\mathcal{A}(#1, #2)}}
\newcommand{\mapassignstruct}[2]{\ensuremath{\mathcal{A}_S(#1, #2)}}
\newcommand{\mapassignarray}[2]{\ensuremath{\mathcal{A}_A(#1, #2)}}
\newcommand{\mapassignmap}[2]{\ensuremath{\mathcal{A}_M(#1, #2)}}
\newcommand{\smtrefcnt}{\ensuremath{\textit{refcnt}}}
\newcommand{\smtnewref}{\ensuremath{\textit{ref}}}
\newcommand{\lhs}{\textit{lhs}\xspace}
\newcommand{\rhs}{\textit{rhs}\xspace}

\newcommand\copyrighttext{%
	\footnotesize \textit{\textbf{Authors' manuscript.}}\quad Published in P. M\"uller (Ed.): \textit{ESOP 2020}, LNCS 12075, 2020. The final publication is available at Springer via {\url{https://doi.org/10.1007/978-3-030-44914-8\_9}}. }
\newcommand\copyrightnotice{%
	\begin{tikzpicture}[remember picture,overlay]
	\node[anchor=south,yshift=2.5cm] at (current page.south) {\fbox{\parbox{\dimexpr\textwidth-\fboxsep-\fboxrule\relax}{\copyrighttext}}};
	\end{tikzpicture}%
}

\begin{document}
\title{SMT-Friendly Formalization of the Solidity Memory Model}
%
%
\author{
\'{A}kos Hajdu%
\inst{1}%
\thanks{The author was also affiliated with SRI International as an intern during this project.
Supported by the ÚNKP-19-3 New National Excellence Program of the Ministry for Innovation and Technology.}%
\and
Dejan Jovanovi\'{c}%
\inst{2}%
}
\authorrunning{\'{A}. Hajdu and D. Jovanovi\'{c}}
%
\institute{
Budapest University of Technology and Economics, Budapest, Hungary\\
\email{hajdua@mit.bme.hu}
\and
SRI International, New York City, USA\\
\email{dejan.jovanovic@sri.com}
}
\maketitle              
\begin{abstract}
Solidity is the dominant programming language for Ethereum smart contracts. This
paper presents a high-level formalization of the Solidity language with a focus
on the memory model. The presented formalization covers all features of the
language related to managing state and memory. In addition, the formalization we
provide is effective: all but few features can be encoded in the quantifier-free
fragment of standard SMT theories. This enables precise and efficient reasoning
about the state of smart contracts written in Solidity. The formalization is
implemented in the \solcverify verifier and we provide an extensive set of tests
that covers the breadth of the required semantics. We also provide an evaluation on
the test set that validates the semantics and shows the novelty of the approach
compared to other Solidity-level contract analysis tools.%
\copyrightnotice%
\end{abstract}


\section{Introduction}

Ethereum \cite{wood2014ethereum} is a public blockchain platform that provides a
novel computing paradigm for developing decentralized applications. Ethereum
allows the deployment of arbitrary programs (termed smart
contracts~\cite{szabo1994smart}) that operate over the blockchain state. The
public can interact with the contracts via transactions. It is currently the most
popular public blockchain with smart contract functionality. While the nodes
participating in the Ethereum network operate a low-level, stack-based virtual
machine (EVM) that executes the compiled smart contracts, the contracts
themselves are mostly written in a high-level, contract-oriented programming
language called Solidity~\cite{soliditydoc}.

Even though smart contracts are generally short, they are no less prone to
errors than software in general. In the Ethereum context, any flaws in the
contract code come with potentially devastating financial consequences (such as
the infamous DAO exploit~\cite{dhillon2017dao}). This has inspired a great
interest in applying formal verification techniques to Ethereum smart contracts (see e.g., \cite{atzei2017survey} or \cite{chen2019survey} for
surveys).
In order to apply formal verification of any kind, be it static analysis or model checking, the first step is to formalize the semantics of the programming language that the smart contracts are written in. Such semantics should not only remain an exercise in formalization, but should preferably be developed, resulting in precise and automated verification tools.

Early approaches to verification of Ethereum smart contracts focused mostly on
formalizing the low-level virtual machine precisely (see, e.g.,
~\cite{bhargavan2016formal,grishchenko2018semantic,hildenbrandt2017kevm,hirai2017defining,amani2018towards}).
However, the unnecessary details of the EVM execution model make it difficult to reason about high-level functional properties of
contracts (as they were written by developers) in an effective and automated way. For Solidity-level properties of smart contracts, Solidity-level semantics are
preferred. While some aspects of Solidity have been studied and
formalized~\cite{jiao2018executable,bartoletti2019minimal,crafa2019solidity,zakrzewski2018towards}, the semantics of the Solidity memory model still lacks a detailed and precise formalization that also enables automation.

The memory model of Solidity has various unusual and non-trivial behaviors,
providing a fertile ground for
potential bugs. Smart contracts have access to two classes of data storage: a
permanent storage that is a part of the global blockchain state, and a transient
local memory used when executing transactions. While the local memory uses a
standard heap of entities with references, the permanent storage has pure value
semantics (although pointers to storage can be declared locally). This memory
model that combines both value and reference semantics, with all interactions
between the two, poses some interesting challenges but also offers great
opportunities for automation. For example, the value semantics of storage
ensures non-aliasing of storage data. This can, if supported by an appropriate
encoding of the semantics, potentially improve both the precision and effectiveness
of reasoning about contract storage.

This paper provides a formalization of the Solidity semantics
in terms of a simple SMT-based intermediate language
that covers all
features related to managing contract storage and memory. A major
contribution of our formalization is that all but few of its elements
can be encoded in the quantifier-free fragment of standard SMT
theories. Additionally, our formalization captures the value semantics of
storage with implicit non-aliasing information of storage entities. This allows
precise and effective verification of Solidity smart contracts using modern SMT
solvers. The formalization is implemented in the open-source \solcverify tool~\cite{vstte2019},
which is a modular verifier for Solidity based on SMT solvers. We validate the
formalization and demonstrate its effectiveness by evaluating it on a
comprehensive set of tests that exercise the memory model. We show that our
formalization significantly improves the precision and soundness compared to existing
Solidity-level verifiers, while remarkably outperforming low-level EVM-based tools in terms of efficiency.

\section{Background}

\subsection{Ethereum}

Ethereum~\cite{wood2014ethereum,antonopoulus2018mastering} is a generic
blockchain-based distributed computing platform. The Ethereum ledger is a
storage layer for a database of accounts (identified by addresses) and the data
associated with the accounts. Every account has an associated balance in Ether
(the native cryptocurrency of Ethereum). In addition, an account can also be
associated with the executable bytecode of a contract and the contract
state.

Although Ethereum contracts are deployed to the blockchain in the form of the
bytecode of the Ethereum Virtual Machine (EVM) \cite{wood2014ethereum}, they are
generally written in a high-level programming language called
Solidity~\cite{soliditydoc} and then compiled to EVM bytecode. After deployment, the
contract is publicly accessible and its code cannot be modified. An external
user, or another contract, can interact with a contract through its API by
invoking its public functions. This can be done by issuing a transaction that
encodes the function to be called with its arguments, and contains the
contract's address as the recipient. The Ethereum network then executes the
transaction by running the contract code in the context of the contract
instance.

A contract instance has access to two different kinds of memory during its
lifetime: contract storage and memory.\footnote{There is an additional data
location named \emph{calldata} that behaves the same as memory, but is used to
store parameters of external functions. For simplicity, we omit it in this paper.}
\emph{Contract storage} is a dedicated data store for a contract to store its
persistent state. At the level of the EVM, it is an array of 256-bit storage
\emph{slots} stored on the blockchain. Contract data that fits into a slot, or
can be sliced into fixed number of slots, is usually allocated starting from
slot 0. More complex data types that do not fit into a fixed number of slots,
such as mappings, or dynamic arrays, are not supported directly by the EVM.
Instead, they are implemented by the Solidity compiler using storage as a hash
table where the structured data is distributed in a deterministic collision-free
manner. \emph{Contract memory} is used during the execution of a transaction on the
contract, and is deleted after the transaction finishes. This is where function
parameters, return values and temporary data can be allocated and stored.

\subsection{Solidity}

Solidity~\cite{soliditydoc} is the high-level programming language supporting
the development of Ethereum smart contracts. It is a full-fledged
object-oriented programming language with many features focusing on enabling
rapid development of Ethereum smart contracts.
The focus of this paper is the semantics of the Solidity memory model: the
Solidity view of contract storage and memory, and the operations that can modify
it. Thus, we restrict the presentation to a generous fragment of Solidity that
is relevant for discussing and formalizing the memory model.
An example contract that
illustrates relevant features is shown in Figure~\ref{fig:ex_storage}, and the
abstract syntax of the targeted fragment is presented in
Figure~\ref{fig:soliditysyntax}.
We omit
parts of Solidity that are not relevant to the memory model (e.g., inheritance,
loops, blockchain-specific members). We also omit low-level, unsafe features
that can break the Solidity memory model abstractions (e.g.,
\codeintext{assembly} and \codeintext{delegatecall}).


\begin{figure*}[t]
\begin{lstlisting}[language=Solidity,escapechar=§,firstline=3,numbers=none,xleftmargin=.05\linewidth,xrightmargin=.05\linewidth]
contract DataStorage {
	struct Record {
		bool set;
		int[] data;
	}

	mapping(address=>Record) private records;

	function append(address at, int d) public {
		Record storage r = records[at];
		r.set = true;
		r.data.push(d);
	}
	function isset(Record storage r) internal view returns (bool s) {
		s = r.set;
	}
	function get(address at) public view returns (int[] memory ret) {
		require(isset(records[at]));
		ret = records[at].data;
	}
}
\end{lstlisting}
\caption{
An example contract illustrating commonly used features of the Solidity
memory model. The contract keeps an association between addresses and data and
allows users to query and append to their data.
}
\label{fig:ex_storage}
\end{figure*}


\begin{figure}[!htb]
\begin{tabular}{lll}
\soltypnam & $::=$ \soladdrtyp\ $|$ \solinttyp\ $|$ \soluinttyp\ $|$ \solbooltyp & Value types\\
& $|$ \solmaptyp{\soltypnam}{\soltypnam} & Mapping\\
& $|$ \soldynarr{\soltypnam}\ $|$ \solfixarr{\soltypnam}{n} & Arrays\\
& $|$ \textit{StructName} & Struct name\\
\textit{DataLoc} & $::=$ \codeintext{storage} $|$ \codeintext{memory} & Data location\\
\sollval & $::=$ \textit{id} & Identifier\\
& $|$ \solexpr\codeintext{.}\textit{id} & Member access\\
& $|$ \solexpr\codeintext{[}\solexpr\codeintext{]} & Index access\\
\solexpr & $::=$ \sollval & Lvalue\\
& $|$ \solexpr\ \codeintext{?} \solexpr \codeintext{:} \solexpr & Conditional\\
& $|$ \codeintext{new} \soldynarr{\soltypnam}\codeintext{(}\solexpr\codeintext{)} & New memory array\\
& $|$ \textit{StructName}\codeintext{(}\solexpr$^*$\codeintext{)} & New memory struct\\
\solstmt & $::=$ \soltypnam\ \textit{DataLoc}? \textit{id} $[$\codeintext{=} \solexpr$]$\codeintext{;} & Local variable declaration\\
& $|$ $(\sollval)^*$ \codeintext{=} $(\solexpr)^*$\codeintext{;} & Assignment (tuples)\\
& $|$ \sollval\codeintext{.push(}\solexpr\codeintext{);} & Push\\
& $|$ \sollval\codeintext{.pop();} & Pop\\
& $|$ \codeintext{delete} \sollval\codeintext{;} & Delete\\
\textit{StructMem} & $::=$ \soltypnam\ \textit{id}\codeintext{;} & Struct member\\
\textit{StructDef} & $::=$ \codeintext{struct} \textit{StructName} \codeintext{\{} \textit{StructMem}$^*$ \codeintext{\}} & Struct definition\\
\textit{StateVar} & $::=$ \soltypnam\ \textit{id}\codeintext{;} & State variable definition\\
\textit{FunPar} & $::=$ \soltypnam\ \textit{DataLoc}? \textit{id} & Function parameter\\
\textit{Fun} & $::=$ \codeintext{function} \textit{id}\codeintext{(}\textit{FunPar}$^*$\codeintext{)} & Function definition\\
& \hspace{1.5em} [\codeintext{returns (}\textit{FunPar}$^*$\codeintext{)}] \codeintext{\{} \solstmt$^*$ \codeintext{\}} & \\
\textit{Constr} & $::=$ \codeintext{constructor}\codeintext{(}\textit{FunPar}$^*$\codeintext{)} \codeintext{\{} \solstmt$^*$ \codeintext{\}} & Constructor definition\\
\textit{Contract} & $::=$ \codeintext{contract} \textit{id} & Contract definition\\
& \hspace{1.5em} \codeintext{\{}\textit{StructDef}$^*$ \textit{StateVar}$^*$ \textit{Constr}? \textit{Fun}$^*$\codeintext{\}} & \\
\end{tabular}
\caption{Syntax of the targeted Solidity fragment.}
\label{fig:soliditysyntax}
\end{figure}

\paragraph{Contracts.}

Solidity contracts are similar to classes in object-oriented programming. A
contract can define any additional types needed, followed by the declaration of the
\emph{state variables} and contract \emph{functions}, including
an optional single \emph{constructor} function.
The contract's state variables define the only persistent data that the contract
instance stores on the blockchain.
The constructor function is only used once, when a new contract instance is
deployed to the blockchain. Other public contract functions can be invoked
arbitrarily by external users through an Ethereum transaction that encodes the
function call data and designates the contract instance as the recipient of the
transaction.

\begin{example}
The contract \codeintext{DataStorage} in Figure~\ref{fig:ex_storage} defines a
struct type \codeintext{Record}. Then it defines the contract storage as a
single state variable \codeintext{records}. Finally three contract functions are
defined \codeintext{append()}, \codeintext{isset()}, and \codeintext{get()}.
Note that a constructor is not defined and, in this case, a default constructor
is provided to initialize the contract state to default values.
\end{example}
Solidity supports further concepts from object-oriented programming, such as
inheritance, function modifiers, and overloading (also covered by our
implementation~\cite{vstte2019}). However, as these are not
relevant for the formalization of the memory model we omit them to simplify our
presentation.

\paragraph{Types.}

Solidity is statically typed and provides two classes of types: \emph{value}
types and \emph{reference} types. Value types include elementary types such as
addresses, integers, and Booleans that are always passed by value. Reference
types, on the other hand, are passed by reference and include structs, arrays
and mappings. A struct consists of a fixed number of members. An array is either
fixed-size or dynamically-sized and besides the elements of
the base type, it also includes a \codeintext{length} field holding
the number of elements. A mapping is an
associative array mapping keys to values. The important caveat is that the table
does not actually store the keys so it is not possible to check if a key is defined in the map.

\begin{example}
The contract in Figure~\ref{fig:ex_storage} uses the following types. The
\codeintext{records} variable is a mapping from addresses to \codeintext{Record}
structures which, in turn, consist of a Boolean value and a dynamically-sized
integer array. It is a common practice to define a struct with a Boolean member
(\codeintext{set}) to indicate that a mapping value has been set. This is
because Solidity mappings do not store keys: any key can be queried, returning a
default value if no value was associated previously.
\end{example}

\paragraph{Data locations for reference types.}

Data of reference types resides in a \emph{data location} that is either
\emph{storage} or \emph{memory}. Storage is the persistent store used for state
variables of the contract. In contrast, memory is used during execution of a
transaction to store function parameters, return values and local variables, and
it is deleted after the transaction finishes.

Semantics of reference types differ fundamentally depending on the data location
that they are stored in. Layout of data in the memory data location resembles the
memory model common in Java-like programming languages: there is a heap
where reference types are allocated and any entity in the heap can contain
values of value types, and \emph{references} to other memory entities. In
contrast, the storage data location treats and stores all entities, including
those of reference types, as \emph{values} with no references involved.
Mixing storage and memory is not possible: the data location of a reference type
is propagated to its elements and members. This means that storage entities
cannot have references to memory entities, and memory entities cannot have
reference types as values. Storage of a contract can be viewed as a single value
with no aliasing possible.


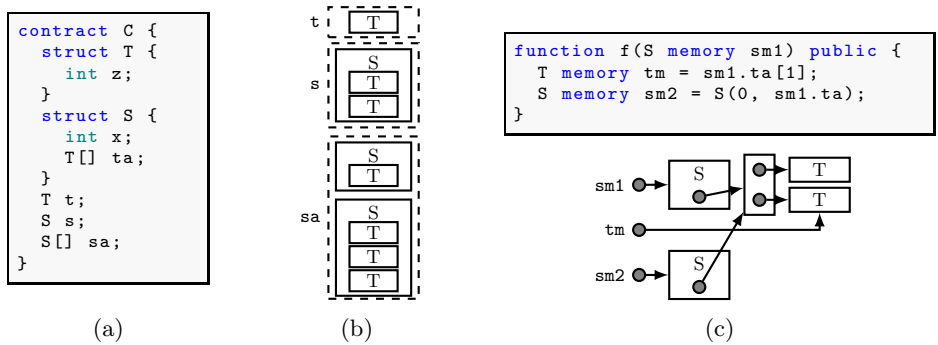
\begin{figure}[htb]
\centering
\begin{subfigure}[b]{.2\linewidth}
\centering
\begin{lstlisting}[language=Solidity,escapechar=§,numbers=none]
contract C {
	struct T {
	  int z;
  }
	struct S {
	  int x;
	  T[] ta;
  }
  T t;
  S s;
  S[] sa;
}
\end{lstlisting}
\caption{}
\label{fig:memory:struct}
\end{subfigure}
\hfill
\begin{subfigure}[b]{.3\linewidth}
\centering
\begin{tikzpicture}[thick,scale=0.8, every node/.style={scale=0.8}]
\tikzstyle{rct}=[rectangle,draw,thick,inner sep=0cm]
\node [rct,dashed,minimum width=1.5cm,minimum height=0.5cm,label=left:\codeintext{t}] at (0,0) {};
\node [rct,dashed,minimum width=1.5cm,minimum height=1.4cm,label=left:\codeintext{s}] at (0,-1.05) {};
\node [rct,dashed,minimum width=1.5cm,minimum height=2.7cm,label=left:\codeintext{sa}] at (0,-3.25) {};

\node [rct,minimum width=0.8cm,minimum height=0.35cm] at (0,0) {T};

\node [rct,minimum width=1.25cm,minimum height=1.2cm] (s1) at (0,-1.05) {};
\node[below] at (s1.north) {S};
\node [rct,minimum width=0.8cm,minimum height=0.35cm] at (0,-1) {T};
\node [rct,minimum width=0.8cm,minimum height=0.35cm] at (0,-1.4) {T};

\node [rct,minimum width=1.25cm,minimum height=0.8cm] (ss0) at (0,-2.4) {};
\node[below] at (ss0.north) {S};
\node [rct,minimum width=0.8cm,minimum height=0.35cm] at (0,-2.55) {T};

\node [rct,minimum width=1.25cm,minimum height=1.6cm] (ss1) at (0,-3.75) {};
\node[below] at (ss1.north) {S};
\node [rct,minimum width=0.8cm,minimum height=0.35cm] at (0,-3.5) {T};
\node [rct,minimum width=0.8cm,minimum height=0.35cm] at (0,-3.9) {T};
\node [rct,minimum width=0.8cm,minimum height=0.35cm] at (0,-4.3) {T};
\end{tikzpicture}
\caption{}
\label{fig:memory:storage}
\end{subfigure}
\hfill
\begin{subfigure}[b]{.45\linewidth}
\centering
\begin{lstlisting}[language=Solidity,escapechar=§,numbers=none]
function f(S memory sm1) public {
	T memory tm = sm1.ta[1];
	S memory sm2 = S(0, sm1.ta);
}
\end{lstlisting}
\begin{tikzpicture}[thick,scale=0.8, every node/.style={scale=0.8}]
\tikzstyle{ptr}=[circle,minimum size=0.2cm,draw,thick,inner sep=0cm,fill=gray]
\tikzstyle{rct}=[rectangle,draw,thick,inner sep=0cm]

\node [ptr,label=left:\codeintext{sm1}] (smptr) at (0,0) {};
\node [rct,minimum width=1cm,minimum height=0.8cm] (sm) at (1,0) {};
\node[below] at (sm.north) {S};
\draw[-latex, thick] (smptr)--(sm);

\node [ptr,label=left:\codeintext{tm}] (tptr) at (0,-0.75) {};

\node [ptr,label=left:\codeintext{sm2}] (sm2ptr) at (0,-1.5) {};
\node [rct,minimum width=1cm,minimum height=0.8cm] (sm2) at (1,-1.5) {};
\node[below] at (sm2.north) {S};
\draw[-latex, thick] (sm2ptr)--(sm2);

\node [rct,minimum width=0.5cm,minimum height=1cm] (tarr) at (2,0) {};
\node [ptr] (tarrptr1) at (1,-0.2) {};
\node [ptr] (tarrptr2) at (1,-1.7) {};
\draw[-latex, thick] (tarrptr1)--(tarr);
\draw[-latex, thick] (tarrptr2)--(tarr);

\node [rct,minimum width=1cm,minimum height=0.4cm] (t1) at (3,0.25) {T};
\node [rct,minimum width=1cm,minimum height=0.4cm] (t2) at (3,-0.25) {T};
\node [ptr] (t1ptr) at (2,0.25) {};
\node [ptr] (t2ptr) at (2,-0.25) {};
\draw[-latex, thick] (t1ptr)--(t1);
\draw[-latex, thick] (t2ptr)--(t2);
\draw[-latex, thick] (tptr)--(3,-0.75)--(t2);

\end{tikzpicture}
\caption{}
\label{fig:memory:memory}
\end{subfigure}
\hfill
\caption{
An example illustrating reference types (structs and arrays) and their layout in storage and memory: (a) a contract defining types and
state variables; (b) an abstract representation of the contract storage as
values; and, (c) a function using the memory data location
and a possible layout of the data in memory.
}
\label{fig:memory}
\end{figure}

\begin{example}
Consider the contract $\codeintext{C}$ defined in
Figure~\ref{fig:memory:struct}. The contract defines two reference
\codeintext{struct} types \codeintext{S} and \codeintext{T}, and  declares state
variables \codeintext{s}, \codeintext{t}, and \codeintext{sa}. These variables
are maintained in storage during the contract lifetime and they are represented
as values with no references within. A potential value of these variables is
shown in Figure~\ref{fig:memory:storage}. On the other hand, the top of
Figure~\ref{fig:memory:memory} shows a function with three variables in the
memory data location, one as the argument to the function, and two defined
within the function. Because they are in memory, these variables are references
to heap locations. Any data of reference types, stored within the structures and
arrays, is also a reference and can be reallocated or assigned to point to an
existing heap location. This means that the layout of the data can contain
arbitrary graphs with arbitrary aliasing. A potential layout of these variables
is shown at the bottom of Figure~\ref{fig:memory:memory}.
\end{example}

\paragraph{Functions.}

Functions are the Solidity equivalent of methods in classes. They receive data as
arguments, perform computations, manipulate state variables and interact
with other Ethereum accounts. Besides accessing the storage of the contract
through its state variables, functions can also define local variables,
including function arguments and return values. Variables of value types are
stored as values on a stack. Variables of reference types must be explicitly declared with
a data location, and are always pointers to an entity in that data
location (storage or memory). A pointer to storage is called a \emph{local
storage pointer}. As the storage is not memory in the usual sense, but a value
instead, one can see storage pointers as encoding a path to one reference
type entity in the storage.

\begin{example}
Consider the example in Figure~\ref{fig:ex_storage}. The local variable
\codeintext{r} in function \codeintext{append()} points to the struct
at index \codeintext{at} of the state variable \codeintext{records}
(residing in the contract storage). In
contrast, the return value \codeintext{ret} of function \codeintext{get()} is a
pointer to an integer array in memory.
\end{example}

\paragraph{Statements and expressions.}

Solidity includes usual programming statements and control structures. To keep
the presentation simple, we focus on the statements that are related to the
formalization of the memory model: local variable declarations, assignments,
array manipulation, and the \codeintext{delete} statement.%
\footnote{Our implementation~\cite{vstte2019} supports a majority of statements, excluding low-level operations (such as inline assembly). Loops are also supported and can be specified with loop invariants.}
Solidity expressions
relevant for the memory model are identifiers, member and array accesses,
conditionals and allocation of new arrays and structs in memory.

If a value is not provided, local variable declarations automatically initialize
the variable to a default value. For reference types in memory, this allocates
new entities on the heap and performs recursive initialization of its members.
For reference types in storage, the local storage pointers must always be
explicitly initialized to point to a storage member. This ensures that no
pointer is ever ``null''. Value types are initialized to their simple default
value (0, false).
Behavior of assignment in Solidity is complex (see Section~\ref{sec:assign})
and depends on the data location of its arguments (e.g., deep copy or pointer assignment).
Dynamically-sized storage arrays can be extended by pushing an element to their
end, or can be shrunk by popping.
The \codeintext{delete} statement assigns the default value (recursively for
reference types) to a given entity based on its type.

\begin{example}
The assignment \codeintext{r.set = true} in the \codeintext{append()} function
of Figure~\ref{fig:ex_storage} is a simple value assignment. On the other hand, \codeintext{ret =
records[at].data} in the \codeintext{get()} function allocates a new array on
the heap and performs a deep copy of data from storage to memory.
\end{example}

\subsection{SMT-Based Programs}


\begin{figure}[htb]
\begin{tabular}{lll}
\textit{TypeName} & $::=$ \textit{int}\ $|$ \textit{bool} & Integer, Boolean\\
& $|$ $[$\textit{TypeName}$]$\textit{TypeName} & SMT array\\
& $|$ \textit{DataTypeName} & SMT datatype\\
\textit{DataTypeDef} & $::=$ \textit{DataTypeName}$((\mathit{id} : \mathit{TypeName})^*)$ & Datatype definition\\
\textit{expr} & $::=$ \textit{id} & Identifier\\
& $|$ $\mathit{expr}[\mathit{expr}]$ & Array read\\
& $|$ $\mathit{expr}[\mathit{expr} \leftarrow \mathit{expr}]$ & Array write\\
& $|$ $\mathit{DataTypeName}(\mathit{expr}^*)$ & Datatype constructor\\
& $|$ $\mathit{expr.id}$ & Member selector\\
& $|$ $ite(\mathit{expr}, \mathit{expr}, \mathit{expr})$ & Conditional\\
& $|$ $\mathit{expr} + \mathit{expr}$ $|$ $\mathit{expr} - \mathit{expr}$ & Arithmetic expression\\
\textit{VarDecl} & $::=$ $\mathit{id} : \mathit{TypeName}$ & Variable declaration\\
\textit{stmt} & $::=$ $\mathit{id} := \mathit{expr}$ & Assignment\\
& $|$ $\mathit{if\ expr\ then\ stmt^*\ else\ stmt^*}$ & If-then-else\\
& $|$ $\mathit{assume}(\mathit{expr})$ & Assumption\\
\textit{Program} & $::=$ $\mathit{DataTypeDef}^* \mathit{VarDecl}^* \mathit{stmt}^*$ & Program definition\\
\end{tabular}
\caption{Syntax of SMT-based programs.}
\label{fig:smtsyntax}
\end{figure}

We formalize the semantics of the Solidity fragment by translating it to a simple
programming language that uses SMT
semantics~\cite{barrett2018satisfiability,biere2009handbook} for the types and
data. The syntax of this language is shown in Figure~\ref{fig:smtsyntax}. The
syntax is purposefully minimal and generic, so that it can be expressed in any
modern SMT-based verification tool (e.g., Boogie~\cite{barnett2006boogie},
Why3~\cite{filliatre2013why3} or Dafny~\cite{leino2010dafny}).\footnote{Our current implementation is based on
Boogie, but we have plans to introduce a generic intermediate representation
that could incorporate alternate backends such as Why3 or Dafny.}

The types of SMT-based programs are the SMT types: simple value types such as
Booleans and mathematical integers, and structured types such as
arrays~\cite{mccarty1962arrays,de2009generalized} and inductive
datatypes~\cite{barrett2007abstract}.
The expressions of the language are standard SMT expressions such as
identifiers, array reads and writes, datatype constructors, member selectors,
conditionals and basic arithmetic~\cite{BarFT-SMTLIB}.
All variables are declared at the beginning of a program. The statements of the language are limited to assignments, the if-then-else statement,
and assumption statement.

SMT-based programs are a good fit for modeling of program semantics. For one, they have clear semantics with no ambiguities. Furthermore, any property of the program can be checked with SMT solvers: the program can be translated directly to a SMT formula by a single static assignment (SSA) transformation.

Note that the syntax requires the left hand side of an assignment to be an
identifier. However, to make our presentation simpler, we will allow array read,
member access and conditional expressions (and their combination) as LHS. Such
constructs can be eliminated iteratively in the following way until only
identifiers appear as LHS in assignments.
\begin{itemize}
	\item $a[i] := e$ is equivalent to $a := a[i \leftarrow e]$.
	\item $d.m_j := e$ is equivalent to $d := D(d.m_1, \ldots, d.m_{j-1}, e, d.m_{j+1}, \ldots, d.m_n)$, where $D$ is the constructor of a datatype with members $m_1, \ldots, m_n$.
	\item $\mathit{ite}(c, t, f) := e$ is equivalent to $\mathit{if}\ c\ \mathit{then}\ t := e\ \mathit{else}\ f := e$.
\end{itemize}


\section{Formalization}

In this section we present our formalization of the Solidity semantics through a
translation that maps Solidity elements to constructs in the SMT-based language.
The formalization is described top-down in separate subsections for types, contracts,
state variables, functions, statements, and expressions.

\subsection{Types}
\label{sec:types}

We use \maptype{.} to denote the function that maps a Solidity type to an SMT
type. This function is used in the translation of contract elements and can, as
a side effect, introduce datatype definitions and variable declarations. This is
denoted with \extradecl{\textit{decl}} in the result of the function. To
simplify the presentation, we assume that such side effects are automatically
added to the preamble of the SMT program. Furthermore, we assume that
declarations with the same name are only added once. We use
\typeof{\textit{expr}} to denote the original (Solidity) type of an expression
(to be used later in the formalization). The definition of \maptype{.} is shown
in Figure~\ref{fig:formalize_types}.


\begin{figure}[thb]
\begin{tabular}{ll}
\maptype{\solbooltyp} & \mappedto \smtbool \\
\maptype{\soladdrtyp} & \mappedto \maptype{\solinttyp}  \mappedto \maptype{\soluinttyp}  \mappedto \smtint \\
\end{tabular}

\tablevspace

\begin{tabular}{ll}
\maptype{\solmaptyp{K}{V}\codeintext{ storage}} & \mappedto \smtarrtyp{\maptype{K}}{\maptype{V}} \\
\maptype{\solmaptyp{K}{V}\codeintext{ storptr}} & \mappedto \smtarrtyp{\smtint}{\smtint} \\
\end{tabular}

\tablevspace

\begin{tabular}{ll}
\maptype{T\codeintext{[}n\codeintext{] storage}} & \mappedto \maptype{T\codeintext{[}\codeintext{] storage}} \\
\maptype{T\codeintext{[}n\codeintext{] storptr}} & \mappedto \maptype{T\codeintext{[}\codeintext{] storptr}} \\
\maptype{T\codeintext{[}n\codeintext{] memory}} & \mappedto \maptype{T\codeintext{[}\codeintext{] memory}} \\
\end{tabular}

\tablevspace

\begin{tabular}{llll}

\maptype{T\codeintext{[}\codeintext{] storage}} & \mappedto $\smtstorarrtype{T}$ & with & $\extradecl{\smtstorarrtype{T}(\mathit{arr} : \smtarrtyp{\smtint}{\maptype{T}}, \mathit{length} : \smtint)}$ \\

\maptype{T\codeintext{[}\codeintext{] storptr}} & \mappedto \smtarrtyp{\smtint}{\smtint} & & \\

\maptype{T\codeintext{[}\codeintext{] memory}} & \mappedto \smtint & with & $\extradecl{\smtmemarrtype{T}(\mathit{arr} : \smtarrtyp{\smtint}{\maptype{T}}, \mathit{length} : \smtint)}$ \\
& & & $\extradecl{\smtarrheap{T} : \smtarrtyp{\smtint}{\smtmemarrtype{T}}}$ \\

\end{tabular}

\tablevspace

\begin{tabular}{llll}
\maptype{\codeintext{struct }S\codeintext{ storage}} & \mappedto $\smtstorstructtype{S}$ & with & $\extradecl{\smtstorstructtype{S}(\ldots, m_i : \maptype{S_i}, \ldots)}$ \\

\maptype{\codeintext{struct }S\codeintext{ storptr}} & \mappedto \smtarrtyp{\smtint}{\smtint} & & \\

\maptype{\codeintext{struct }S\codeintext{ memory}} & \mappedto \smtint & with & $\extradecl{\smtmemstructtype{S}(\ldots, m_i : \maptype{S_i}, \ldots)}$ \\
& & & $\extradecl{\smtstructheap{S} : \smtarrtyp{\smtint}{\smtmemstructtype{S}}}$ \\

\end{tabular}
\caption{
Formalization of Solidity types. Members of \codeintext{struct }$S$ are denoted as $m_i$ with types $S_i$.
}
\label{fig:formalize_types}
\end{figure}

\paragraph{Value types.}

Booleans are mapped to SMT Booleans while other value types are mapped to
SMT integers. Addresses are also mapped to SMT integers so that arithmetic comparison
and conversions between integers and addresses is supported. For
simplicity, we map all integers (signed or unsigned) to SMT
integers.\footnote{Note that this does not capture the precise machine integer semantics, but this is not relevant from the perspective of the memory model.
Precise computation can be provided by relying on SMT bitvectors or modular arithmetic (see, e.g., \cite{vstte2019}).}
Solidity also allows function types to store, pass around, and call functions, but this is not yet supported by our encoding.

\paragraph{Reference types.}

The Solidity syntax does not always require the data location for
variable and parameter declarations. However, for reference types it is always
required (enforced by the compiler), except for state variables that are
always implicitly storage. In our formalization, we assume that the data
location of reference types is a part of the type. As discussed before, memory
entities are always accessed through pointers. However, for storage we
distinguish whether it is the storage reference itself (e.g., state variable) or a
storage pointer (e.g., local variable, function parameter). We denote the former
with \codeintext{storage} and the latter with \codeintext{storptr} in the type
name.
Our modeling of reference types relies on the generalized theory of arrays
\cite{de2009generalized} and the theory of inductive
data-types~\cite{barrett2007abstract}, both of which are supported by modern SMT
solvers (e.g., \cvc \cite{barrett2011cvc4} and \zzz \cite{de2008z3}).

\paragraph{Mappings and arrays.}

%

For both arrays and mappings, we abstract away the implementation details of Solidity and model them with the SMT theory of arrays and
inductive datatypes. We formalize Solidity mappings simply as SMT arrays. Both fixed- and dynamically-sized arrays are translated
using the same SMT type and we only treat them differently in the context of
statements and expressions.
Strings and byte arrays are not discussed here, but we support them as
particular instances of the array type.
To ensure that array size is properly modeled we
keep track of it in the datatype (\textit{length}) along with the actual
elements (\textit{arr}).

For \emph{storage array types} with base type $T$, we introduce an SMT datatype
\smtstorarrtype{T} with a constructor that takes two arguments: an inner SMT
array (\textit{arr}) associating integer indexes and the recursively translated base
type (\maptype{T}), and an integer \textit{length}. The advantage of this
encoding is that the value semantics of storage data is provided by
construction: each array element is a separate entity (no aliasing) and
assigning storage arrays in SMT makes a deep copy. This encoding also
generalizes if the base type is a reference type.

For \emph{memory array types} with base type $T$, we introduce a separate
datatype \smtmemarrtype{T} (side effect). However, memory arrays are stored with
pointer values. Therefore the memory array type is mapped to integers, and a
heap (\smtarrheap{T}) is introduced to associate integers (pointers) with the
actual memory array datatypes. Note that mixing data locations within a reference
type is not possible: the element type of the array has the same data location
as the array itself. Therefore, it is enough to introduce two datatypes per
element type $T$: one for storage and one for memory. In the former case the
element type will have value semantics whereas in the latter case elements will
be stored as pointers.

\paragraph{Structs.}

For each \emph{storage struct type} $S$ the translation introduces an inductive
datatype \smtstorstructtype{S}, including a constructor for each
struct member with types mapped recursively. Similarly to arrays, this
ensures the value semantics of storage such as non-aliasing and deep copy
assignments. For each \emph{memory struct $S$} we also introduce a datatype
\smtmemstructtype{S} and a constructor for each member.%
\footnote{Mappings in Solidity cannot reside in memory. If a struct
	defines a mapping member and it is stored in memory, the mapping is simply
	inaccessible. Such members could be omitted from the constructor.}
However, the memory struct type itself is mapped to integers (pointer) and a heap
(\smtstructheap{S}) is introduced to associate the pointers with the actual
memory struct datatypes. Note that if a memory struct has members with reference types,
they are also pointers, which is ensured recursively by our encoding.

\subsection{Local Storage Pointers}

An interesting aspect of the storage data location is that, although the stored
data has value semantics, it is still possible to define pointers to an entity
in storage within a local context, e.g., with function parameters or local
variables. These pointers are called \emph{local storage pointers}.

\begin{example}
In the \codeintext{append()} function of Figure~\ref{fig:ex_storage} the
variable \codeintext{r} is defined to be a convenience pointer into the storage
map \codeintext{records[at]}. Similarly, the \codeintext{isset()} function takes
a storage pointer to a \codeintext{Record} entity in storage as an argument.
\end{example}
Since our formalization uses SMT datatypes to encode the contract data in storage, it
is not possible to encode these pointers directly. A partial solution would be
to substitute each occurrence of the local pointer with the expression that is
assigned to it when it was defined. However, this approach is too simplistic and
has limitations. Local storage pointers can be reassigned, or assigned
conditionally, or it might not be known at compile
time which definition should be used.
Furthermore, local storage pointers can also be passed
in as function arguments: they can point to different
storage entities for different calls.


\begin{figure}[htb]
\centering
\begin{subfigure}[b]{.16\linewidth}
\centering
\begin{lstlisting}[language=Solidity,escapechar=§,numbers=none]
contract C {
	struct T {
		int z;
	}
	struct S {
		int x;
		T   t;
		T[] ts;
	}
	T   t1;
	S   s1;
	S[] ss;
}
\end{lstlisting}
\caption{}
\label{fig:ex_pack1}
\end{subfigure}
\hfill
\begin{subfigure}[b]{.5\linewidth}
\centering
\begin{tikzpicture}
\tikzstyle{crc}=[rectangle,draw,thick,inner sep=0.1cm, minimum width=0.6cm]

\node [crc] (c) at (0,0) {\scriptsize C};

\node [crc] (t1) at (1.5,0) {\scriptsize T};
\draw[-latex, thick] (c)--(t1) node[midway,above]{\scriptsize t1 (0)};

\node [crc] (s1) at (1.5,-0.7) {\scriptsize S};
\draw[-latex, thick] (c)--(0,-0.7)--(s1) node[midway,above]{\scriptsize s1 (1)};

\node [crc] (s1t) at (3,-0.7) {\scriptsize T};
\draw[-latex, thick] (s1)--(s1t) node[midway,above]{\scriptsize t (0)};

\node [crc] (s1ts) at (3,-1.4) {\scriptsize T[]};
\draw[-latex, thick] (s1)--(1.5,-1.4)--(s1ts) node[midway,above]{\scriptsize ts (1)};

\node [crc] (s1tsi) at (4,-1.4) {\scriptsize T};
\draw[-latex, thick] (s1ts)--(s1tsi) node[midway,above]{\scriptsize (i)};

\node [crc] (ss) at (1.5,-2.1) {\scriptsize S[]};
\draw[-latex, thick] (c)--(0,-2.1)--(ss) node[midway,above]{\scriptsize ss (2)};

\node [crc] (ssi) at (2.5,-2.1) {\scriptsize S};
\draw[-latex, thick] (ss)--(ssi) node[midway,above]{\scriptsize (i)};

\node [crc] (ssit) at (4,-2.1) {\scriptsize T};
\draw[-latex, thick] (ssi)--(ssit) node[midway,above]{\scriptsize t (0)};

\node [crc] (ssits) at (4,-2.8) {\scriptsize T[]};
\draw[-latex, thick] (ssi)--(2.5,-2.8)--(ssits) node[midway,above]{\scriptsize ts (1)};

\node [crc] (ssitsi) at (5,-2.8) {\scriptsize T};
\draw[-latex, thick] (ssits)--(ssitsi) node[midway,above]{\scriptsize (i)};
\end{tikzpicture}
\caption{}
\label{fig:ex_pack2}
\end{subfigure}
\hfill
\begin{subfigure}[b]{.3\linewidth}
\centering
\begin{tabular}{l}
$\unpackptr{\textit{ptr}} =$\\
\hspace{1em}$\textit{ite}(\textit{ptr}[0] = 0,$\\
\hspace{1.5em}$\textit{t1},$\\
\hspace{1.5em}$\textit{ite}(\textit{ptr}[0] = 1,$\\
\hspace{2em}$\textit{ite}(\textit{ptr}[1] = 0,$\\
\hspace{2.5em}$\textit{s1.t},$\\
\hspace{2.5em}$\textit{s1.ts}[\textit{ptr}[2]]),$\\
\hspace{2em}$\textit{ite}(\textit{ptr}[2] = 0,$\\
\hspace{2.5em}$\textit{ss}[\textit{ptr}[1]].t,$\\
\hspace{2.5em}$\textit{ss}[\textit{ptr}[1]].\textit{ts}[\textit{ptr}[3]])))$\\
\end{tabular}
\caption{}
\label{fig:ex_pack3}
\end{subfigure}
\caption{
An example of packing and unpacking: (a) contract with struct definitions and
state variables; (b) the storage tree of the contract for type \codeintext{T};
and (c) the unpacking expression for storage pointers of type \codeintext{T}.
}
\label{fig:ex_pack}
\end{figure}
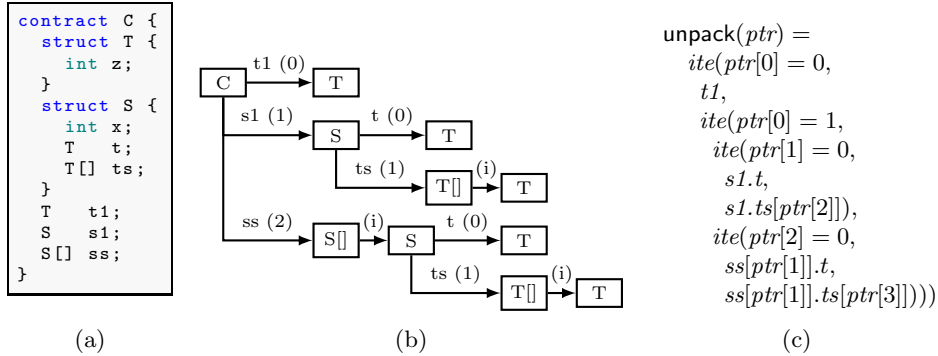

We propose an approach to encode local storage pointers while overcoming these
limitations. Our encoding relies on the fact that storage data of a contract
can be viewed as a finite-depth tree of values. As such, each element of the
stored data can be uniquely identified by a finite path leading to it.%
\footnote{Solidity does support a limited form of recursive data-types. Such types
could make the storage a tree of potentially arbitrary depth. We chose not to
support such types as recursion is non-existing in Solidity types used in
practice.}

\begin{example}
Consider the contract \codeintext{C} in Figure~\ref{fig:ex_pack1}. The contract
defines structs \codeintext{T} and \codeintext{S}, and state variables of these
types. If we are interested in all storage entities of type
\codeintext{T}, we can consider the sub-tree of the contract storage tree that
has leaves of type \codeintext{T}, as depicted in Figure~\ref{fig:ex_pack2}.
The root of the tree is the contract itself, with indexed sub-nodes for
state variables, in order. For nodes of struct type there are indexed sub-nodes
leading to its members, in order. For each node of array type there is a sub-node
for the base type. Every pointer to a storage \codeintext{T} entity can be
identified by a path in this tree: by fixing the index to each state variable,
member, and array index, as seen in brackets in Figure~\ref{fig:ex_pack2}, such
paths can be encoded as an array of integers. For example, the state variable
\codeintext{t1} can be represented as $[0]$, the member \codeintext{s1.t}
as $[1, 0]$, and \codeintext{ss[8].ts[5]} as $[2, 8, 1, 5]$.
\end{example}
This idea allows us to encode storage pointer types (pointing to arrays, structs
or mappings) simply as SMT arrays (\smtarrtyp{\smtint}{\smtint}). The novelty of
our approach is that storage pointers can be encoded and passed around, while
maintaining the value semantics of storage data, without the need for
quantifiers to describe non-aliasing. To encode storage pointers, we need to
address initialization and dereference of storage pointers,
while assignment is simply an assignment of array values. When a storage pointer
is initialized to a concrete expression, we \emph{pack} the indexed path to the
storage entity (that the expression references) into an array value. When a storage
pointer is dereferenced (e.g., by indexing into or accessing a member), the
array is \emph{unpacked} into a conditional expression that will evaluate to a
storage entity by decoding paths in the tree.

\paragraph{Storage tree.}

The storage tree for a given type $T$ can be easily obtained by filtering
the AST nodes of the contract definition to only include state variable
declarations and to, further, only include nodes that lead to a sub-node of type
$T$. We denote the storage tree for type $T$ as $\storagetree(T)$.%
\footnote{In our implementation we do not explicitly compute the storage tree
but instead traverse directly the AST provided by the Solidity compiler.}

\paragraph{Packing.}


\begin{figure}[htb]
\begin{algorithm}[H]
	\SetAlgoLined
	\SetKwProg{Fn}{def}{:}{}
	\Fn{\packpath($\textit{node}, \textit{subExprs}, d, \textit{result}$)}{
		\ForEach{\textit{expr} in \textit{subExprs}}{
  		\If{$\solexpr = \textit{id} \vee \solexpr = e.\textit{id}$}{
	  		find edge $\textit{node} \xrightarrow{\textit{id}\;(i)} \textit{child}$\;
			  $\textit{result} := \textit{result}[d \leftarrow i]$\;
	  	}
	  	\If{$\solexpr = e[\textit{idx}]$}{
		  	find edge $\textit{node} \xrightarrow{(i)} \textit{child}$\;
		  	$\textit{result} := \textit{result}[d \leftarrow \mapexpr{\textit{idx}}]$\;
		  }
			$\mathit{node}, d := \mathit{child}, d+1$\;
		}
		\Return{$\mathit{result}$}
	}
	\Fn{\packptr{\solexpr}}{
	  $\mathit{baseExprs} := $ list of base sub-expressions of $\solexpr$\;
		$\mathit{baseExpr} := \mathsf{car}(\mathit{baseExprs})$\;
		\If{\textit{baseExpr} is a state variable}{
			\Return{$\packpath(\storagetree(\typeof{\solexpr}), \mathit{baseExprs}, 0, \mathit{\smtconstarr{\smtint}{\smtint}{0}})$}
		}
		\If{\textit{baseExpr} is a storage pointer}{
  		$\mathit{result} := \smtconstarr{\smtint}{\smtint}{0}$\;
			$\mathit{prefix} := \mapexpr{\mathit{baseExpr}}$\;
			\ForEach{path to a leaf in $\storagetree(\typeof{\mathit{baseExpr}})$}{
				$\mathit{pathResult}, \mathit{pathCond} := \mathit{prefix}, \mathit{true}$\;
				\ForEach{$k$th edge on the path with label id (i)}{
					$\mathit{pathCond} := \mathit{pathCond} \wedge \mathit{prefix}[k] = i$
				}
				$\mathit{pathResult} := \packpath(\mathit{leaf}, \mathsf{cdr}(\mathit{baseExprs}), \mathit{len}(\mathit{path}), \mathit{pathResult})$\;
				$\mathit{result} := \mathit{ite}(\mathit{pathCond}, \mathit{pathResult}, \mathit{result})$\;
			}
			\Return{$\mathit{result}$}
		}
	}
\end{algorithm}
\caption{Packing of an expressions. It returns a symbolic array expression that, when evaluated, can identify the path to the storage entity that the expression references.}
\label{fig:formalize_pack}
\end{figure}

Given an expression (such as \codeintext{ss[8].ts[5]}), \packptr{.} uses
the storage tree for the type of the expression and encodes it to an array (e.g., $[2, 8, 1, 5]$) by
fitting the expression into the tree.
Pseudocode for \packptr{.} is shown in Figure~\ref{fig:formalize_pack}.
To start, the expression is decomposed into a list of base sub-expressions. The base expression of an identifier \textit{id} is \textit{id} itself. For an array index $e[i]$ or a member access $e.m_i$ it is recursively the base expressions of $e$. We call the first element of this list (denoted by $\mathsf{car}$) the base expression (the innermost base expression). The base expression is always either a state variable or a storage pointer, and we consider these two cases separately.

If the \emph{base expression is a state variable}, we simply align the
expression along the storage tree with the \packpath function. The \packpath
function takes the list of base sub-expressions, and the storage tree to use for
alignment, and then processes the expressions in order. If the current
expression is an identifier (state variable or member access), the algorithm
finds the outgoing edge annotated with the identifier (from the current node)
and writes the index into the result array. If the expression is an index
access, the algorithm maps and writes the index expression (symbolically) in the
array. The expression mapping function \mapexpr{.} is introduced later in
Section~\ref{sec:expr}.

If the \emph{base expression is a storage pointer}, the process is more general since the ``start'' of the packing must accommodate any point in storage where the base expression can point to.
In this case the algorithm finds all paths
to leaves in the tree of the base pointer, identifies the condition for taking
that path and writes the labels on the path to an array. Then it uses \packpath
to continue writing the array with the rest of the expression (denoted by $\mathsf{cdr}$), as before. Finally, a
conditional expression is constructed with all the conditions and packed arrays.
Note, that the type of this conditional is still an SMT array of integers as it
is the case for a single path.

\begin{example}
For contract in Figure~\ref{fig:ex_pack1},
\packptr{\codeintext{ss[8].ts[5]}} produces $[2, 8, 1, 5]$ by calling \packpath on the base sub-expressions $[\codeintext{ss}, \codeintext{ss[8]}, \codeintext{ss[8].ts}, \codeintext{ss[8].ts[5]}]$.
First, $2$ is added as \codeintext{ss} is the state variable with index 2.
Then, \codeintext{ss[8]} is an index access so \codeintext{8} is mapped to $8$ and added to the result.
Next, \codeintext{ss[8].ts} is a member access with \codeintext{ts} having the index $1$.
Finally, \codeintext{ss[8].ts[5]} is an index access so \codeintext{5} is mapped to $5$ and added.
\end{example}

\paragraph{Unpacking.}


\begin{figure}[htb]
\begin{algorithm}[H]
	\SetAlgoLined
	\SetKwProg{Fn}{def}{:}{}
	\Fn{\unpackptr{\textit{ptr}}}{
		\Return{$\textit{unpack}(ptr, \storagetree(\typeof{\textit{ptr}}), \textit{empty}, 0)$}\;
	}
	\Fn{\unpackptr{\textit{ptr}, \textit{node}, \textit{expr}, d}}{

		$\textit{result} := \textit{empty}$\;

		\lIf{\textit{node} has no outgoing edges}{\textit{result} := \textit{expr}}

		\If{\textit{node} is contract}{
			\ForEach{edge $\textit{node} \xrightarrow{\textit{id}\;(i)} \textit{child}$}{
				$\textit{result} := \textit{ite}(\textit{ptr}[\textit{d}] = i, \unpackptr{\textit{ptr}, \textit{child}, id, \textit{d} + 1}, \textit{result})$\;
			}
		}

		\If{\textit{node} is struct}{
			\ForEach{edge $\textit{node} \xrightarrow{\textit{id}\;(i)} \textit{child}$}{
				$\textit{result} := ite(\textit{ptr}[\textit{d}] = i, \unpackptr{\textit{ptr}, \textit{child}, \textit{expr}.id, \textit{d} + 1), result}$\;
			}
		}

		\If{\textit{node} is array/mapping with edge $\textit{node} \xrightarrow{(i)} \textit{child}$}{
			\textit{result} := $\unpackptr{\textit{ptr}, \textit{child}, \textit{expr}[\textit{ptr}[\textit{d}]], \textit{d} + 1}$\;
		}

		\Return{\textit{result}}\;
	}
	\vspace{0.5em}
\end{algorithm}
\caption{Unpacking of a local storage pointer into a conditional expression.}
\label{fig:formalize_unpack}
\end{figure}

The opposite of \packptr{} is \unpackptr{}, shown in Figure~\ref{fig:formalize_unpack}. This function takes a storage
pointer (of type \smtarrtyp{\smtint}{\smtint}) and produces a conditional
expression that decodes any given path into one of the leaves of the storage
tree.
The function recursively traverses the tree starting from the contract node and
accumulates the expressions leading to the leaves. The function creates
conditionals when branching, and when a leaf is reached the accumulated
expression is simply returned. For contracts we process edges corresponding to
each state variable by setting the subexpression to be the state variable
itself. For structs we process edges corresponding to each member by wrapping
the subexpression into a member access. For both contracts and structs, the
subexpressions are collected into a conditional as separate cases. For arrays
and mappings we process the single outgoing edge by wrapping the subexpression
into an index access using the current element (at index $d$) of the pointer.

\begin{example}
For example, the conditional expression corresponding to the tree in
Figure~\ref{fig:ex_pack2} can be seen in Figure~\ref{fig:ex_pack3}. Given a
pointer $\textit{ptr}$, if $\textit{ptr}[0] = 0$ then the conditional evaluates
to \textit{t1}. Otherwise, if $\textit{ptr}[0] = 1$ then \textit{s1} has to be
taken, where two leaves are possible: if $\textit{ptr}[1] = 0$ then the result
is \textit{s1.t} otherwise it is \textit{s1.ts}$[\textit{ptr}[2]]$, and so on.
If \textit{ptr} is $[2, 8, 1, 5]$ then the conditional evaluates exactly to
\codeintext{ss[8].ts[5]} from which \textit{ptr} was packed.%
\footnote{Note that due to the ``else'' branches, unpack is a is a non-injective
surjective function. For example, $[a, 8, 1, 5]$ with any $a \geq 2$ would
evaluate to the same slot. However this does not affect our encoding as
pointers cannot be compared and pack always returns the same (unique) values.}
\end{example}

Note that with inheritance and libraries~\cite{soliditydoc} it is possible that
a contract defines a type $T$ but has no nodes in its storage tree. The contract
can still define functions with storage pointers to $T$, which can be called by
derived contracts that define state variables of type $T$. In such cases we
declare an array of type \smtarrtyp{\smtint}{\maptype{T}}, called the \emph{default context}, and
unpack storage pointers to $T$ as if the default context was a state variable.
This allows us to reason about abstract contracts and libraries, modeling that
their storage pointers can point to arbitrary entities not yet declared.

\subsection{Contracts, State Variables, Functions}

The focus of our discussion is the Solidity memory model and, for presentation
purposes, we assume a minimalist setting where the important aspects of
storage and memory can be presented: we assume a single contract and a single
function to translate. Interactions between multiple functions are handled
differently depending on the verification approach. For example, in modular
verification functions are checked individually against specifications (pre- and
post-conditions) and function calls are replaced by their
specification~\cite{vstte2019}.

\paragraph{State variables.}

Each state variable $s_i$ of a contract is mapped to a variable declaration
$s_i: \maptype{\typeof{s_i}}$ in the SMT program.%
\footnote{%
	Generalizing this to multiple
	contracts can be done directly by using a separate one-dimensional heap for each
	state variable, indexed by a receiver parameter ($\textit{this}:
	\textit{address}$) identifying the current contract instance (see, e.g.,
	\cite{vstte2019}).%
}
The data location of state variables is always storage. As
discussed previously, reference types are mapped using SMT datatypes and arrays,
which ensures non-aliasing by construction. While Solidity optionally allows inline
initializer expressions for state variables, without the loss of generality we
can assume that they are initialized in the constructor using regular
assignments.

\paragraph{Functions calls.}

From the perspective of the memory model, the only important aspect of function
calls is the way parameters are passed in and how function return
values are treated. Our formalization is general in that it allows us to treat
both of the above as plain assignments (explained later in Section~\ref{sec:assign}). For each parameter
$p_i$ and return value $r_i$ of a function, we add
declarations $p_i: \maptype{\typeof{p_i}}$ and $r_i: \maptype{\typeof{r_i}}$ in the SMT program.
Note that for reference types appearing as parameters or return values of the
function, their types are either memory or storage pointers.

\paragraph{Memory allocation.}

In order to model allocation of new memory entities, while keeping some
non-aliasing information, we introduce an allocation counter $\smtrefcnt:
\smtint$ variable in the preamble of the SMT program. This counter is
incremented for each allocation of memory entities and used as the address of
the new entity. For each parameter $p_i$ with memory data location we include an
assumption $\textit{assume}(p_i \leq \smtrefcnt)$ as they can be arbitrary
pointers, but should not alias with new allocations within the function. Note
that if a parameter of memory pointer type is a reference type containing other
references, such non-aliasing constraints need to be assumed recursively~\cite{leino1997ecstatic}.
This can be done for structs by enumerating members. But, for dynamic arrays it
requires quantification that is nevertheless still decidable (array property
fragment \cite{bradley2006whats}).

\paragraph{Initialization and default values.}


\begin{figure}[t]
\begin{tabular}{llll}
\defval{\solbooltyp} & \mappedto \textit{false} \\
\defval{\soladdrtyp} & \mappedto \defval{\solinttyp} & \mappedto \defval{\soluinttyp} & \mappedto $0$ \\
\end{tabular}

\tablevspace

\begin{tabular}{ll}
\defval{\solmaptyp{K}{V}} & \mappedto \smtconstarr{\maptype{K}}{\maptype{V}}{\defval{V}} \\
\end{tabular}

\tablevspace

\begin{tabular}{ll}
\defval{T\codeintext{[}\codeintext{] storage}} & \mappedto \defval{T\codeintext{[}0\codeintext{] storage}} \\
\defval{T\codeintext{[}\codeintext{] memory}} & \mappedto \defval{T\codeintext{[0}\codeintext{] memory}} \\
\end{tabular}

\tablevspace

\begin{tabular}{llll}
\defval{T\codeintext{[}n\codeintext{] storage}} & \mappedto & $\smtstorarrtype{T}(\smtconstarr{\smtint}{\maptype{T}}{\defval{T}}, n)$ & \\

\defval{T\codeintext{[}n\codeintext{] memory}} & \mappedto & \extradecl{\smtnewref\ : \smtint} (fresh symbol) \\
& & $\extrastmt{\smtnewref := \smtrefcnt := \smtrefcnt + 1}$ & \\
& & $\extrastmt{\smtarrheap{T}[\smtnewref].\textit{length} := n}$ & \\
& & $\extrastmt{\smtarrheap{T}[\smtnewref].\textit{arr}[i] := \defval{T}}$ & for $0 \leq i \leq n$\\
& & $\smtnewref$ & \\
\end{tabular}

\tablevspace

\begin{tabular}{llll}
\defval{\codeintext{struct }S\codeintext{ storage}} & \mappedto & $\textit{StorStruct}_S(\ldots, \defval{S_i}, \ldots)$ & \\
\defval{\codeintext{struct }S\codeintext{ memory}} & \mappedto & \extradecl{\smtnewref\ : \smtint} (fresh symbol) \\
& & $\extrastmt{\smtnewref := \smtrefcnt := \smtrefcnt + 1}$ & \\
& & $\extrastmt{\smtstructheap{S}[\smtnewref].m_i = \defval{S_i}}$ & for each $m_i$\\
& & $\smtnewref$ & \\

\end{tabular}
\caption{Formalization of default values. We denote \codeintext{struct }$S$ members as $m_i$ with types $S_i$.}
\label{fig:formalize_defval}
\end{figure}

If we are translating the constructor function, each state variable $s_i$ is
first initialized to its default value with a statement $s_i :=
\defval{\typeof{s_i}}$. For regular functions, we set
each return value $r_i$ to its default value with a statement $r_i := \defval{\typeof{r_i}}$. We use \defval{.}, as defined in
Figure~\ref{fig:formalize_defval}, to denote the function that maps a Solidity
type to its default value as an SMT expression. Note that, as a side effect,
this function can do allocations for memory entities, introducing extra declarations and statements,
denoted by \extradecl{\textit{decl}} and \extrastmt{\textit{stmt}}.
As expected, the default value is $\textit{false}$ for Booleans and $0$ for
other primitives that map to integers. For mappings from $K$ to $V$, the default
value is an SMT constant array returning the default value of the value type $V$ for each key $k \in K$
(see, e.g., \cite{de2009generalized}). The default value of storage arrays is
the corresponding datatype value constructed with a constant array of
the default value for base type $T$, and a length of $n$ or $0$ for fixed- or dynamically-sized
arrays. For storage structs, the default value is the corresponding datatype
value constructed with the default values of each member.

The default value of uninitialized
memory pointers is unusual. Since Solidity doesn't support ``null'' pointers, a
new entity is automatically allocated in memory and initialized to default
values (which might include additional recursive initialization).
Note, that for fixed-size arrays Solidity enforces that the array size $n$ must
be an integer literal or a compile time constant, so setting each element to its
default value is possible without loops or quantifiers. Similarly for structs,
each member is recursively initialized, which is again possible by explicitly
enumerating each member.

\subsection{Statements}
\label{sec:stmt}

We use \mapstmt{.} to denote the function that translates Solidity statements to
a list of statements in the SMT program. It relies on the type mapping function
\maptype{.} (presented previously in Section~\ref{sec:types}) and on the
expression mapping function \mapexpr{.} (to be introduced in
Section~\ref{sec:expr}). Furthermore, we define a helper function
\mapassign{.}{.} dedicated to modeling Solidity assignments (to be discussed in Section~\ref{sec:assign}).

The definition of \mapstmt{.} is shown in Figure~\ref{fig:formalize_stmt}. As a
side effect, extra declarations can be introduced to the preamble of the SMT
program (denoted by \extradecl{\textit{decl}}).
The Solidity documentation~\cite{soliditydoc} does not precisely state the order of evaluating subexpressions in statements.
It only specifies that subnodes are processed before the parent node.
This problem is independent form the discussion of the memory models so we assume that side effects of subexpressions are added in the same order as it is implemented in the compiler.
Furthermore, if a subexpression is mapped multiple times, we assume that the side effects are only added once.
This makes our presentation simpler by introducing fewer temporary variables.


\begin{figure}[htb]
\begin{tabular}{lll}

\mapstmt{\textit{T}\ \textit{id}} &  \mappedto $\extradecl{\textit{id} : \maptype{T}}$; & $\mapassign{\textit{id}}{\defval{T}}$\\

\mapstmt{\textit{T}\ \textit{id} = \textit{expr}} & \mappedto $\extradecl{\textit{id} : \maptype{T}}$; & $\mapassign{\textit{id}}{ \mapexpr{\textit{expr}}}$ \\

\mapstmt{\codeintext{delete }\textit{e}} & \mappedto & $\mapassign{\mapexpr{e}}{\defval{\typeof{e}}}$ \\

\end{tabular}

\tablevspace

\begin{tabular}{lll}
\mapstmt{l_1, \ldots, l_n \codeintext{ = } r_1, \ldots, r_n} \mappedto & $\extradecl{\textit{tmp}_i : \maptype{\typeof{r_i}}}$ & for $1 \leq i \leq n$ (fresh symbols) \\
& $\mapassign{\textit{tmp}_i}{\mapexpr{r_i}}$ & for $1 \leq i \leq n$\\
& $\mapassign{\mapexpr{l_i}}{\textit{tmp}_i}$ & for $n \geq i \geq 1$ (reversed)\\
\end{tabular}

\tablevspace

\begin{tabular}{lll}

\mapstmt{e_1\codeintext{.push(}e_2\codeintext{)}} & \mappedto &
$\mapassign{\mapexpr{e_1}.\textit{arr}[\mapexpr{e_1}.\textit{length}]}{\mapexpr{e_2}}$\\
& & $\mapexpr{e_1}.\textit{length} := \mapexpr{e_1}.\textit{length} + 1$ \\

\mapstmt{e\codeintext{.pop()}} & \mappedto & $\mapexpr{e}.\textit{length} := \mapexpr{e}.\textit{length} - 1$ \\
& & $\mapassign{\mapexpr{e}.\textit{arr}[\mapexpr{e}.\textit{length}]}{\defval{\elementtypeof{\mapexpr{e}}}}$\\

\end{tabular}

\tablevspace

\caption{Formalization of statements.}
\label{fig:formalize_stmt}
\end{figure}

Local variable declarations introduce a variable declaration with the same
identifier in the SMT program by mapping the type.%
\footnote{Without the loss of generality we assume that identifiers in Solidity
are unique. The compiler handles scoping and assigns an unique identifier to
each declaration.}
If an initialization expression is given, it is mapped using \mapexpr{.} and
assigned to the variable. Otherwise, the default value is used as defined by
\defval{.} in Figure~\ref{fig:formalize_defval}.
Delete assigns the default value for a type, which is simply mapped to an
assignment in our formalization.
Solidity supports multiple assignments as one statement with a tuple-like
syntax. The documentation~\cite{soliditydoc} does not specify the behavior
precisely, but the compiler first evaluates the RHS and LHS tuples (in this
order) from left to right and then assignment is performed component-wise
from right to left.


\begin{figure}[t]
\begin{minipage}[t]{.59\linewidth}
\begin{lstlisting}[language=Solidity,escapechar=§,numbers=none]
contract C {
	struct S { int x; }
	
	S s1, s2, s3;
	
	function primitiveAssign() {
		s1.x = 1; s2.x = 2; s3.x = 3;
		(s1.x, s3.x, s2.x) = (s3.x, s2.x, s1.x);
		// s1.x == 3, s2.x == 1, s3.x == 2
	}
	function storageAssign() {
		s1.x = 1; s2.x = 2; s3.x = 3;
		(s1, s3, s2) = (s3, s2, s1);
		// s1.x, s2.x, s3.x are all equal to 1
	}
}
\end{lstlisting}
\caption{Example illustrating the right-to-left assignment order and the treatment of reference types in storage in tuple assignment.}
\label{fig:ex_crazyswap}
\end{minipage}
\hfill
\begin{minipage}[t]{.35\linewidth}
\begin{lstlisting}[language=Solidity,escapechar=§,numbers=none]
contract C {
	struct S { int x; }
	
	S[] a;
	
	constructor() {
		a.push(S(1));
		S storage s = a[0];
		a.pop();
		assert(s.x == 1); // Ok
		// Following is error
		// assert(a[0].x == 1);
	}
}
\end{lstlisting}
\caption{Example illustrating a dangling pointer to storage.}
\label{fig:ex_dangling}
\end{minipage}
\end{figure}

\begin{example}
Consider the tuple assignment in function \codeintext{primitiveAssign()} in
Figure~\ref{fig:ex_crazyswap}. From right to left, \codeintext{s2.x} is assigned
first with the value of \codeintext{s1.x} which is $1$. Afterwards, when
\codeintext{s3.x} is assigned with \codeintext{s2.x}, the already evaluated (old) value of
$2$ is used instead of the new value $1$. Finally, \codeintext{s1.x} gets the
old value of \codeintext{s3.x}, i.e., $3$. Note however, that storage
expressions on the RHS evaluate to storage pointers. Consider, for example, the function
\codeintext{storageAssign()} in Figure~\ref{fig:ex_crazyswap}. From right to left,
\codeintext{s2} is assigned first, with a pointer to \codeintext{s1} making
\codeintext{s2.x} become $1$. However, as opposed to primitive types, when
\codeintext{s3} is assigned next, \codeintext{s2} on the RHS is a storage
pointer and thus the new value in the storage of \codeintext{s2} is assigned to
\codeintext{s3} making \codeintext{s3.x} become $1$. Similarly,
\codeintext{s1.x} also becomes $1$ as the new value behind \codeintext{s3} is used.
\end{example}

Array push increases the length and assigns the given expression as the last
element. Array pop decreases the length and sets the removed element to its
default value. While the removed element can no longer be accessed via indexing
into an array (a runtime error occurs), it can still be accessed via local
storage pointers (see Figure~\ref{fig:ex_dangling}).%
\footnote{The current version (0.5.x) of Solidity supports resizing arrays by assigning to the length member. However, this behavior is dangerous and has been since removed in the next version (0.6.0) (see \url{https://solidity.readthedocs.io/en/v0.6.0/060-breaking-changes.html}). Therefore, we do not support this in our encoding.}

\subsection{Assignments}
\label{sec:assign}

Assignments between reference types in Solidity can be either pointer
assignments or value assignments, involving deep copying and possible new
allocations in the latter case. We use \mapassign{\lhs}{\rhs} to denote the function that assigns
a \rhs SMT expression to a \lhs SMT expression based on their original types and data
locations. The definition of \mapassign{.}{.} is shown in
Figure~\ref{fig:formalize_assign}. Value type assignments are simply mapped to an SMT
assignment. To make our presentation more clear, we subdivide the other cases
into separate functions for array, struct and mapping operands, denoted by
\mapassignarray{.}{.}, \mapassignstruct{.}{.} and \mapassignmap{.}{.}
respectively.


\begin{figure}[htb]
\begin{tabular}{lll}
\mapassign{\lhs}{\rhs} \mappedto & $\lhs := \rhs$ & for value type operands\\
\mapassign{\lhs}{\rhs} \mappedto & \mapassignmap{\lhs}{\rhs} & for mapping type operands\\
\mapassign{\lhs}{\rhs} \mappedto & \mapassignstruct{\lhs}{\rhs} & for struct type operands\\
\mapassign{\lhs}{\rhs} \mappedto & \mapassignarray{\lhs}{\rhs} & for array type operands\\
\end{tabular}

\tablevspace

\begin{tabular}{lll}
\mapassignmap{\lhs: \codeintext{sp}}{\rhs: \codeintext{s}} & \mappedto $\lhs := \packptr{\rhs}$ & \\
\mapassignmap{\lhs: \codeintext{sp}}{\rhs: \codeintext{sp}} & \mappedto $\lhs := \rhs$ & \\
\mapassignmap{\lhs}{\rhs} & \mappedto \{\} & (all other cases) \\
\end{tabular}

\tablevspace

\begin{tabular}{lll}
\mapassignstruct{\lhs: \codeintext{s}}{\rhs: \codeintext{s}} & \mappedto & $\lhs := \rhs$\\
\mapassignstruct{\lhs: \codeintext{s}}{\rhs: \codeintext{m}} & \mappedto & \mapassign{\lhs.m_i}{\smtstructheap{\typeof{\rhs}}[\rhs].m_i} for each $m_i$ \\
\mapassignstruct{\lhs: \codeintext{s}}{\rhs: \codeintext{sp}} & \mappedto & \mapassignstruct{\lhs}{\unpackptr{\rhs}} \\
\mapassignstruct{\lhs: \codeintext{m}}{\rhs: \codeintext{m}} & \mappedto & $\lhs := \rhs$ \\
\mapassignstruct{\lhs: \codeintext{m}}{\rhs: \codeintext{s}} & \mappedto & $\lhs := \smtrefcnt := \smtrefcnt + 1$ \\
& & \mapassign{\smtstructheap{\typeof{\lhs}}[\lhs].m_i}{\rhs.m_i} for each $m_i$ \\
\mapassignstruct{\lhs: \codeintext{m}}{\rhs: \codeintext{sp}} & \mappedto & \mapassignstruct{\lhs}{\unpackptr{\rhs}} \\
\mapassignstruct{\lhs: \codeintext{sp}}{\rhs: \codeintext{s}} & \mappedto & $\lhs := \packptr{\rhs}$ \\
\mapassignstruct{\lhs: \codeintext{sp}}{\rhs: \codeintext{sp}} & \mappedto & $\lhs := \rhs$ \\
\end{tabular}

\tablevspace

\begin{tabular}{lll}
\mapassignarray{\lhs: \codeintext{s}}{\rhs: \codeintext{s}} & \mappedto & $\lhs := \rhs$ \\
\mapassignarray{\lhs: \codeintext{s}}{\rhs: \codeintext{m}} & \mappedto & $\lhs := \smtarrheap{\typeof{\rhs}}[\rhs]$ \\
\mapassignarray{\lhs: \codeintext{s}}{\rhs: \codeintext{sp}} & \mappedto & $\mapassignarray{\lhs}{\unpackptr{\rhs}}$ \\
\mapassignarray{\lhs: \codeintext{m}}{\rhs: \codeintext{m}} & \mappedto & $\lhs := \rhs$ \\
\mapassignarray{\lhs: \codeintext{m}}{\rhs: \codeintext{s}} & \mappedto & $\lhs := \smtrefcnt := \smtrefcnt + 1$\\
& & $\smtarrheap{\typeof{\lhs}}[\lhs] := \rhs$ \\
\mapassignarray{\lhs: \codeintext{m}}{\rhs: \codeintext{sp}} & \mappedto & $\mapassignarray{\lhs}{\unpackptr{\rhs}}$ \\
\mapassignarray{\lhs: \codeintext{sp}}{\rhs: \codeintext{s}} & \mappedto & $\lhs := \packptr{\rhs}$ \\
\mapassignarray{\lhs: \codeintext{sp}}{\rhs: \codeintext{sp}} & \mappedto & $\lhs := \rhs$\\
\end{tabular}

\caption{Formalization of assignment based on different type categories and data locations for the LHS and RHS. We use \codeintext{s}, \codeintext{sp} and \codeintext{m} after the arguments to denote storage, storage pointer and memory types respectively.}
\label{fig:formalize_assign}
\end{figure}

\paragraph{Mappings.}

As discussed previously, Solidity prohibits direct assignment of mappings.
However, it is possible to declare a storage pointer to a mapping, in which case
the RHS expression is packed. It is also possible to assign two storage
pointers, which simply assigns pointers. Other cases are a no-op.%
\footnote{This is consequence of the fact that keys are not stored in mappings and so the assignment is impossible to perform.}


\begin{figure}[htb]
\centering
\begin{tabular}{|l|ccc|}
\hline
lhs/rhs   & Storage        & Memory         & Stor.ptr.      \\\hline
Storage   & Deep copy      & Deep copy      & Deep copy      \\
Memory    & Deep copy      & Pointer assign & Deep copy      \\
Stor.ptr. & Pointer assign & Error          & Pointer assign \\\hline
\end{tabular}
\caption{Semantics of assignment between array and struct operands based on their data location.}
\label{fig:copying}
\end{figure}

\paragraph{Structs and arrays.}

For structs and arrays the semantics of assignment is summarized in Figure~\ref{fig:copying}. However, there are some notable details in various cases that we expand on below.

Assigning anything \emph{to storage} LHS always causes a deep copy.
If the RHS is storage, this is simply mapped to a datatype assignment in our encoding (with an additional unpacking if the RHS is storage pointer).%
\footnote{This also causes mappings to be copied, which contradicts the
	current semantics. However, we chose to keep the deep copy as
	assignments of mappings is planned to be disallowed in the future
	(see \url{https://github.com/ethereum/solidity/issues/7739}).
}
If the RHS is memory, deep copy for structs can be done member
wise by accessing the heap with the RHS pointer and
performing the assignment recursively (as members can be reference types
themselves). For arrays, we access the datatype corresponding to the array via
the heap and do an assignment, which does a deep copy in SMT. Note however, that
this only works if the base type of the array is a value type. For reference types,
memory array elements are pointers and would require being dereferenced during
assignment to storage. As opposed to struct members, the number of array
elements is not known at compile time so loops or quantifiers have to be used
(as in traditional software analysis). However, this is a special case, which
can be encoded in the decidable array property fragment~\cite{bradley2006whats}.
Assigning storage (or storage pointer) \emph{to memory} is also a deep copy but in the other direction. However, instead overwriting the existing memory entity,
a new one is allocated (recursively for reference typed elements or members). We model this by incrementing the reference counter,
storing it in the LHS and then accessing the heap for deep copy using the new
pointer.

\subsection{Expressions}
\label{sec:expr}

\begin{figure}[!h]
\begin{tabular}{l}
\mapexpr{\mathit{id}} \mappedto $\mathit{id}$ \\
\end{tabular}

\tablevspace

\begin{tabular}{lll}
\mapexpr{\solexpr\codeintext{.}\textit{id}} \mappedto &
$\mapexpr{\solexpr}.\mapexpr{\textit{id}}$ &
if \typeof{\solexpr} $=$ \codeintext{struct} $S$ \codeintext{storage} \\

\mapexpr{\solexpr\codeintext{.}\textit{id}} \mappedto &
\unpackptr{\mapexpr{\solexpr}}$.$\mapexpr{\textit{id}} &
if \typeof{\solexpr} $=$ \codeintext{struct} $S$ \codeintext{storptr} \\

\mapexpr{\solexpr\codeintext{.}\textit{id}} \mappedto &
$\smtstructheap{S}[\mapexpr{\solexpr}].\mapexpr{\textit{id}}$ &
if \typeof{\solexpr} $=$ \codeintext{struct} $S$ \codeintext{memory} \\

\mapexpr{\solexpr\codeintext{.}\textit{id}} \mappedto &
$\mapexpr{\solexpr}.\mapexpr{\textit{id}}$ &
if \typeof{\solexpr} $= T$\codeintext{[] storage} \\

\mapexpr{\solexpr\codeintext{.}\textit{id}} \mappedto &
\unpackptr{\mapexpr{\solexpr}}$.$\mapexpr{\textit{id}} &
if \typeof{\solexpr} $= T$\codeintext{[] storptr}\\

\mapexpr{\solexpr\codeintext{.}\textit{id}} \mappedto &
$\smtarrheap{T}[\mapexpr{\solexpr}].\mapexpr{\textit{id}}$ &
if \typeof{\solexpr} $= T$\codeintext{[] memory} \\
\end{tabular}

\tablevspace

\begin{tabular}{lll}
\mapexpr{\solexpr\codeintext{[}\textit{idx}\codeintext{]}} \mappedto &
$\mapexpr{\solexpr}.\mathit{arr}[\mapexpr{\textit{idx}}]$ &
if \typeof{\solexpr} $= T$\codeintext{[] storage}\\

\mapexpr{\solexpr\codeintext{[}\textit{idx}\codeintext{]}} \mappedto &
$\unpackptr{\mapexpr{\solexpr}}.\mathit{arr}[\mapexpr{\textit{idx}}]$ &
if \typeof{\solexpr} $= T$\codeintext{[] storptr} \\

\mapexpr{\solexpr\codeintext{[}\textit{idx}\codeintext{]}} \mappedto &
$\smtarrheap{T}[\mapexpr{\solexpr}].\mathit{arr}[\mapexpr{\textit{idx}}]$ &
if \typeof{\solexpr} $= T$\codeintext{[] memory} \\

\mapexpr{\solexpr\codeintext{[}\textit{idx}\codeintext{]}} \mappedto &
$\mapexpr{\solexpr}[\mapexpr{\textit{idx}}]$ &
if \typeof{\solexpr} $=$ \codeintext{mapping(}$K$\codeintext{=>}$V$\codeintext{) storage}\\

\mapexpr{\solexpr\codeintext{[}\textit{idx}\codeintext{]}} \mappedto &
$\unpackptr{\mapexpr{\solexpr}}[\mapexpr{\textit{idx}}]$ &
if \typeof{\solexpr} $=$ \codeintext{mapping(}$K$\codeintext{=>}$V$\codeintext{) storptr} \\
\end{tabular}

\tablevspace

\begin{tabular}{ll}
\mapexpr{\textit{cond}\codeintext{ ? }\solexpr_T\codeintext{ : }\solexpr_F} \mappedto & \extradecl{$\textit{var}_T$ : \maptype{\typeof{\textit{cond}\codeintext{ ? }\solexpr_T\codeintext{ : }\solexpr_F}}} (fresh symbol) \\
& \extradecl{$\textit{var}_F$ : \maptype{\typeof{\textit{cond}\codeintext{ ? }\solexpr_T\codeintext{ : }\solexpr_F}}} (fresh symbol) \\
& \extrastmt{\mapassign{\textit{var}_T}{\mapexpr{\solexpr_T}}} \\
& \extrastmt{\mapassign{\textit{var}_F}{\mapexpr{\solexpr_F}}} \\
& $\textit{ite}(\mapexpr{\textit{cond}}, \textit{var}_T, \textit{var}_F)$ \\
\end{tabular}

\tablevspace

\begin{tabular}{lll}
\mapexpr{\codeintext{new }\textit{T}\codeintext{[](}\textit{expr}\codeintext{)}} \mappedto & \extradecl{\smtnewref\ : \smtint} (fresh symbol) & \\
& \extrastmt{$\smtnewref := \smtrefcnt := \smtrefcnt + 1$} & \\
& \extrastmt{$\smtarrheap{T}[\smtnewref].\textit{length} := \mapexpr{\textit{expr}}$} & \\
& \extrastmt{$\smtarrheap{T}[\smtnewref].\textit{arr}[i] := \defval{T}$} & for $0 \leq i \leq \mapexpr{\textit{expr}}$ \\
& $\smtnewref$ & \\
\end{tabular}

\tablevspace

\begin{tabular}{lll}
\mapexpr{\textit{S}\codeintext{(}\ldots, \textit{expr}_i, \ldots\codeintext{)}} \mappedto & \extradecl{\smtnewref\ : \smtint} (fresh symbol) & \\
& \extrastmt{$\smtnewref := \smtrefcnt := \smtrefcnt + 1$} & \\
&\extrastmt{$\smtstructheap{S}[\smtnewref].m_i := \mapexpr{\textit{expr}_i}$} & for each member $m_i$ \\
& $\smtnewref$ & \\
\end{tabular}
\caption{Formalization of expressions. We denote \codeintext{struct }$S$ members as $m_i$ with types $S_i$.}
\label{fig:formalize_expr}
\end{figure}

We use \mapexpr{.} to denote the function that translates a Solidity expression
to an SMT expression. As a side effect, declarations and statements might be
introduced (denoted by \extradecl{\textit{decl}} and \extrastmt{\textit{stmt}}
respectively). The definition of
\mapexpr{.} is shown in Figure~\ref{fig:formalize_expr}.
As discussed in Section~\ref{sec:stmt} we assume that side effects are added from subexpressions in the proper order and only once.

Member access is mapped to an SMT member access by mapping the base expression
and the member name. There is an extra unpacking step for storage pointers and a heap access for memory. Note that the only valid member for arrays is \codeintext{length}.
Index access is mapped to an SMT array read by mapping the base
expression and the index, and adding en extra member access for arrays to get the
inner array \textit{arr} of elements from the datatype.
Furthermore, similarly to member accesses, an extra unpacking step is needed for storage pointers and a heap access for memory.

Conditionals in Solidity can be mapped to an SMT conditional in
general. However, data locations can be different for the true and false
branches, causing possible side effects.
Therefore, we first introduce fresh variables for the true and false branch with the
common type (of the whole conditional), then make assignments using \mapassign{.}{.} and finally use the new variables in the conditional.
The documentation~\cite{soliditydoc} does not specify the common type, but the compiler returns memory if any of the branches is memory, and storage pointer otherwise.

Allocating a new array in memory increments the reference counter, sets the
length and the default values for each element (recursively). Note that
in general the length
might not be a compile time constant,in which case setting default values could be encoded with the array
property fragment (similarly to deep copy in
assignments)~\cite{bradley2006whats}.
Allocating a new memory struct also increments the reference counter and sets each value by translating the provided arguments.


\section{Evaluation}

The formalization described in this paper serves as the basis of our Solidity
verification tool \solcverify~\cite{vstte2019}.%
\footnote{\solcverify is open source, available at \url{https://github.com/SRI-CSL/solidity}.
Besides certain low-level constructs (such as inline assembly) \solcverify supports a majority of Solidity features that we omitted from the presentation, including inheritance, function modifiers, for/while loops and if-then-else.}
In this section we provide an
evaluation of the presented formalization and our implementation by validating
it on a set of relevant test cases. For illustrative purposes we also compare
our tool with other available Solidity analysis tools.\footnote{All tests, with a Truffle test harness, a docker container with all the tools, and all individual results are available at \url{https://github.com/dddejan/solidity-semantics-tests}.}


``Real world'' contracts currently deployed on Ethereum (e.g., contract
available on Etherscan) have limited value for evaluating memory
model semantics. Many such contracts use old compiler versions with constructs
that are not supported anymore, and do not use newer features. There are also
many toy and trivial contracts that are deployed but not used, and popular
contracts (e.g. tokens) are over-represented with many duplicates. Furthermore,
the inconsistent usage of \codeintext{assert} and
\codeintext{require}~\cite{vstte2019} makes evaluation hard. Evaluating the
memory semantics requires contracts that exercise diverse features of the memory
model. There are larger dApps that do use more complex features (e.g., Augur or
ENS), but these contracts also depend on many other features (e.g. inheritance,
modifiers, loops) that would skew the results.

Therefore we have manually developed a set of tests that try to capture the
interesting behaviors and corner cases of the Solidity memory semantics. The
tests are targeted examples that do not use irrelevant features. The set is
structured so that every target test behavior is represented with a test case
that sets up the state, exercises a specific feature and checks the correctness
of the behavior with assertions. This way a test should only pass if the tool
provides a correct verification result by modeling the targeted feature
precisely. The correctness of the tests themselves is determined by running them
through the EVM with no assertion failures. Test cases are expanded to use all
reference types and combinations of reference types. This includes structures,
mappings, dynamic and fixed-size arrays, both single- and multi-dimensional.

The tests are organized into the following classes.
Tests in the \textsf{assignment} class check whether the assign statement is
properly modeled. This includes assignments in the same data location, but also
assignments across data locations that need deep copying, and assignments and
re-assignments of memory and storage pointers.
The \textsf{delete} class of tests checks whether the \codeintext{delete}
statement is properly modeled.
Tests in the \textsf{init} class check whether variable and data initialization
is properly modeled. For variables in storage, we check if they are properly
initialized to default values in the contract constructor. Similarly, we check
whether memory variables are properly initialized to provided values, or default
values when no initializer is provided.
The \textsf{storage} class of tests checks whether storage itself is properly modeled for various reference types, including for example non-aliasing.
Tests in the \textsf{storageptr} class check whether storage pointers are
modeled properly. This includes checking if the model properly treats storage
pointers to various reference types, including nested types. In addition, the
tests check that the storage pointers can be properly passed to functions and
ensure non-aliasing for distinct parts of storage.


For illustrative purposes we include a comparison with the following available
Solidity analysis tools: \myth v0.21.17~\cite{mueller2018smashing}, \verisol
v0.1.1-alpha~\cite{lahiri2018formal}, and \smtchecker v0.5.12~\cite{alt2018smt}.
\myth is a Solidity symbolic execution tool
that runs analysis at the level of the EVM bytecode. \verisol is similar to \solcverify
in that it uses Boogie to model the Solidity contracts, but takes the
traditional approach to modeling memory and storage with pointers and
quantifiers. \smtchecker is an SMT-based analysis module built into the Solidity
compiler itself. There are other tools that can be found in the literature, but
they are either basic prototypes that cannot handle realistic features we are
considering, or are not available for direct comparison.

\begin{table}[htb]
	\caption{Results of evaluating \myth, \verisol, \smtchecker, and \solcverify on our test suite.}
	\label{tbl:results}
	\begin{center}
		\begin{tabular}{| l | c  c  c  c | r |}
\hline
\rowcolor[gray]{.95}
\textsf{assignment} (102) & correct & incorrect & unsupported & timeout & time (s) \\
\hline
\myth & 94 & 0 & 0 & 8 & 1655.14 \\ 
\verisol & 10 & 61 & 31 & 0 & 175.27 \\ 
\smtchecker & 6 & 9 & 87 & 0 & 15.25 \\ 
\solcverify & 78 & 8 & 16 & 0 & 62.81 \\ 

\hline
\rowcolor[gray]{.95}
\textsf{delete} (14) & correct & incorrect & unsupported & timeout & time (s) \\
\hline
\myth & 13 & 1 & 0 & 0 & 47.51 \\ 
\verisol & 3 & 8 & 3 & 0 & 24.66 \\ 
\smtchecker & 0 & 0 & 14 & 0 &  0.30 \\ 
\solcverify & 7 & 1 & 6 & 0 &  9.02 \\ 

\hline
\rowcolor[gray]{.95}
\textsf{init} (18) & correct & incorrect & unsupported & timeout & time (s) \\
\hline
\myth & 15 & 3 & 0 & 0 & 59.67 \\ 
\verisol & 7 & 8 & 3 & 0 & 28.82 \\ 
\smtchecker & 0 & 0 & 18 & 0 &  0.41 \\ 
\solcverify & 13 & 5 & 0 & 0 & 11.88 \\ 

\hline
\rowcolor[gray]{.95}
\textsf{storage} (27) & correct & incorrect & unsupported & timeout & time (s) \\
\hline
\myth & 27 & 0 & 0 & 0 & 310.40 \\ 
\verisol & 12 & 15 & 0 & 0 & 43.45 \\ 
\smtchecker & 2 & 0 & 25 & 0 &  1.32 \\ 
\solcverify & 27 & 0 & 0 & 0 & 17.61 \\ 

\hline
\rowcolor[gray]{.95}
\textsf{storageptr} (164) & correct & incorrect & unsupported & timeout & time (s) \\
\hline
\myth & 164 & 0 & 0 & 0 & 1520.29 \\ 
\verisol & 128 & 19 & 17 & 0 & 203.93 \\ 
\smtchecker & 4 & 18 & 142 & 0 & 21.93 \\ 
\solcverify & 164 & 0 & 0 & 0 & 96.92 \\ 

\hline
\end{tabular}

	\end{center}
\end{table}
%
%
We ran the experiments on a machine with Intel Xeon E5-4627 v2 @ 3.30GHz CPU enforcing a 60s timeout and a memory limit of 64GB. Results are shown
in Table~\ref{tbl:results}. As expected, \myth has the most consistent results
on our test set. This is because \myth models contract semantics at the EVM
level and does not need to model complex Solidity semantics. Nevertheless, the
results also indicate that the performance penalty for this precision is
significant (8 timeouts). \verisol, as the closest to our approach, still
doesn't support many features and has a significant amount of false reports for
features that it does support. Many false reports are because their model
of storage is based on pointers and tries to ensure storage consistency with
the use of quantifiers. \smtchecker doesn't yet
support the majority of the Solidity features that our tests target.

Based on the results, \solcverify performs well on our test set, matching the precision of \myth at
very low computational cost. The few false alarms we have are either due to
Solidity features that we chose to not implement (e.g., proper treatment of
\codeintext{mapping} assignments), or parts of the semantics that we only
implemented partially (such as deep copy of arrays with reference types and recursively initializing memory objects). There are no technical difficulties in supporting them and they are planned in the future.


\section{Related Work}

There is a strong push in the Ethereum community to apply formal methods to
smart contract verification. This includes many attempts to formalize the
semantics of smart contracts, both at the level of EVM and Solidity.

\paragraph{EVM-level semantics.}

Bhargavan et al.~\cite{bhargavan2016formal} decompile a fragment of EVM
to F*, modeling EVM as a stack based machine with word and byte arrays for
storage and memory.
Grishchenko et al.~\cite{grishchenko2018semantic} extend this work by providing
a small step semantics for EVM.
\textsc{Kevm}~\cite{hildenbrandt2017kevm} provides an executable formal
semantics of EVM in the K framework.
Hirai~\cite{hirai2017defining} formalizes EVM in Lem, a language used by some
interactive theorem provers.
Amani et al.~\cite{amani2018towards} extends this work by defining a program
logic to reason about EVM bytecode.

\paragraph{Solidity-level semantics.}

Jiao et al.~\cite{jiao2018executable} formalize the operational
semantics of Solidity in the K framework. Their formalization focuses on the
details of bit-precise sizes of types, alignment and padding in storage. They
encode storage slots, arrays and mappings with the full encoding of hashing.
However, the formalization does not describe assignments (e.g., deep copy) apart from simple cases. Furthermore, user defined structs are also not
mentioned. In contrast, our semantics is high-level and abstracts away some
details (e.g., hashes, alignments) to enable efficient verification.
Additionally, we provide proper modeling of different cases for
assignments between storage and memory.
Bartotelli et al.~\cite{bartoletti2019minimal} propose \textsc{TinySol}, a
minimal core calculus for a subset of Solidity, required to model basic
features such as asset transfer and reentrancy. Contract data is modeled
as a key value store, with no differences in storage and
memory, or in value and reference types.
Crafa et al.~\cite{crafa2019solidity} introduce Featherweight Solidity, a
calculus formalizing core features of the language, with focus on primitive types.
Data locations and
reference types are not discussed, only mappings are mentioned briefly.
The main focus is on the type system and type
checking. They propose an improved type system that can statically detect unsafe
casts and callbacks.
The closest to our work is the work of
Zakrzewski~\cite{zakrzewski2018towards}, a Coq formalization focusing on
functions, modifiers, and the memory model. The memory model is treated
similarly: storage is a mapping from names to storage objects (values), memory
is a mapping from references to memory objects (containing references
recursively) and storage pointers define a path in storage. Their formalization
is also high-level, without considering alignment, padding or hashing. The
formalization is provided as big step functional semantics in Coq. While the
paper presents some example rules, the formalization does not cover all cases.
For example the details of assignments (e.g., memory to storage), push/pop for
arrays, treating memory aliasing and new expressions. Furthermore, our approach
focuses on SMT and modular verification, which enables automated reasoning.


\section{Conclusion}

We presented a high-level SMT-based formalization of the Solidity memory model
semantics. Our formalization covers all aspects of the language related to
managing both the persistent contract storage and the transient local memory.
The novel encoding of storage pointers as arrays allows us to precisely model
non-aliasing and deep copy assignments between storage entities without the need
for quantifiers. The memory model forms the basis of our Solidity-level modular
verification tool \solcverify. We developed a suite of test cases exercising all
aspects of memory management with different combinations of reference types.
Results indicate that our memory model outperforms
existing Solidity-level tools in terms of soundness and precision, and is on par with
low-level EVM-based implementations, while having a significantly lower computational cost for discharging verification conditions.

\bibliographystyle{splncs04}
\bibliography{references}

\end{document}